\documentclass[11pt,a4paper]{article}
\usepackage{jheppub}
\usepackage{comment}
\usepackage{tikz}
\usetikzlibrary{decorations.pathmorphing,decorations.markings}

\tikzset{
  psi/.style={
    decoration={
      markings,
      mark=at position 0.53 with {\arrow{>}}
    },
    postaction={decorate},
    double,
    double distance=1pt
  },
  psiNoArrow/.style={
    decoration={
      markings,
      mark=at position 0.6 with 
    },
    postaction={decorate},
    double,
    double distance=1pt
  },
  nucleon/.style={
    decoration={
      markings,
      mark=at position 0.6 with {\arrow{>}}
    },
    postaction={decorate}
  },
  external/.style={},  
  gluon/.style={
  decorate, draw=black, 
  decoration={coil,amplitude=4pt, segment length=5pt}
  },
  particle/.style={draw=black, postaction={decorate}, decoration={markings,mark=at position .5 with {\arrow[draw=black]{>}}}},
 photon/.style={decorate, decoration={snake,amplitude=2pt, segment length=5pt}, draw=black}
}

\newcommand{\diagramOne}{
  \begin{tikzpicture}[baseline=-3pt]
    \draw[particle] (-1.5, 0) -- (0, 0);
    \draw[particle] (1.5, 0) -- (0, 0);
    
    \draw[psi]  (0,1.4)   arc (90:270:0.7cm) ;
    \draw[particle]  (0,1.4)   arc (90:-90:0.7cm) ;

    \draw[photon] (0,1.4) -- (0,2.4);
    
    \node at (-1.6, 0.3) [external]{$\mu_L$};
    \node at (1.6, 0.3) [external]{$\mu_R$};
    \node at (0, 2.6) [external]{$\gamma$};
    \node at (-1.1, 1) [external]{$\psi_{3/2}$};
    \node at (1, 1) [external]{$\nu$};
    \node at (-1, 2.3) [external]{a)};

    \fill[black] (0,1.4) circle (0.06cm);
    \fill[black] (0,0) circle (0.06cm);

  \end{tikzpicture}
}

\newcommand{\diagramThree}{
  \begin{tikzpicture}[baseline=-3pt]
    \draw[particle] (-2, 0) -- (-1, 0);
    \draw[psi] (1, 0) -- (-1, 0);
    \draw[particle] (1, 0) -- (1.7, 0);
    \draw[particle] (2.4, 0) -- (1.7, 0);
    
    \draw (1.7-0.1,0+0.1) -- (1.7+0.1,0-0.1);
    \draw (1.7+0.1,0+0.1) -- (1.7-0.1,0-0.1);
    
    \draw[photon] (-1, 0) -- (0, 1.41);
    \draw[photon] (1, 0) -- (0, 1.41);
    
    \draw[photon] (0, 1.41) -- (0, 2.41);
    
    \fill[black] (-1,0) circle (0.06cm);
    \fill[black] (1,0) circle (0.06cm);
    \fill[black] (0,1.41) circle (0.06cm);
    
    \node at (-2.1, 0.3) [external]{$\mu_L$};
    \node at (2.5, 0.3) [external]{$\mu_R$};

    \node at (0, 2.6) [external]{$\gamma$};
    \node at (0, -0.4) [external]{$\psi_{3/2}$};
    \node at (-0.8, 0.8) [external]{$W$};
    \node at (0.8, 0.8) [external]{$W$};
    \node at (-1, 2.3) [external]{b)};

  \end{tikzpicture}
}

\newcommand{\diagramFive}{
  \begin{tikzpicture}[baseline=-3pt]
    \draw[particle] (-2, 0) -- (-1, 0);
    \draw[photon] (1, 0) -- (-1, 0);
    \draw[particle] (1, 0) -- (1.7, 0);
    \draw[particle] (2.4, 0) -- (1.7, 0);
    
    \draw (1.7-0.1,0+0.1) -- (1.7+0.1,0-0.1);
    \draw (1.7+0.1,0+0.1) -- (1.7-0.1,0-0.1);
    
    \draw[psi] (0,1.41) -- (-1, 0);
    \draw[particle] (0,1.41) -- (1,0);
    
    \draw[photon] (0, 1.41) -- (0, 2.41);
    
    \fill[black] (-1,0) circle (0.06cm);
    \fill[black] (1,0) circle (0.06cm);
    \fill[black] (0,1.41) circle (0.06cm);
    
    \node at (-2.1, 0.3) [external]{$\mu_L$};
    \node at (2.5, 0.3) [external]{$\mu_R$};

    \node at (0, 2.6) [external]{$\gamma$};
    \node at (0, -0.4) [external]{$W$};
    \node at (-1., 0.8) [external]{$\psi_{3/2}$};
    \node at (0.8, 0.8) [external]{$\nu$};
    \node at (-1, 2.3) [external]{c)};

  \end{tikzpicture}
}

\usepackage{hyperref}
\usepackage{amsmath,amsfonts,amssymb}
\usepackage{booktabs}
\usepackage{enumitem}
\usepackage{graphicx}
\usepackage{siunitx}
\usepackage{xcolor}
\usepackage{verbatim}
\setlist[itemize]{leftmargin=*}

\newcommand{\be}{\begin{equation}}
\newcommand{\ee}{\end{equation}}
\newcommand{\bea}{\begin{equation}\begin{aligned}}
\newcommand{\eea}{\end{aligned}\end{equation}}
\newcommand{\td}{{\rm d}}

\newcommand{\GeV}{{\rm GeV}}
\newcommand{\TeV}{{\rm TeV}}

\newcommand{\eg}{{\it e.g.}}

\newcommand{\gsim}{\lower.7ex\hbox{$\;\stackrel{\textstyle>}{\sim}\;$}}
\newcommand{\lsim}{\lower.7ex\hbox{$\;\stackrel{\textstyle<}{\sim}\;$}}

\definecolor{grey}{cmyk}{0,0,0,0.75}
\definecolor{tangerine}{cmyk}{0,0.5,1,0}
\definecolor{darkgreen}{cmyk}{1,0,1,0.23}
\definecolor{Red}{rgb}{1,0,0}
\definecolor{Blue}{rgb}{0,0,1}
\definecolor{Green}{rgb}{0,1,0}
\definecolor{Grey}{cmyk}{0,0,0,0.75}
\definecolor{Tangerine}{cmyk}{0,0.5,1,0}
\definecolor{Darkgreen}{cmyk}{1,0,1,0.23}
\definecolor{Cyan}{cmyk}{1,0,0,0}
\definecolor{Yellow}{cmyk}{0,0,1,0}
\definecolor{darkblue}{cmyk}{1,0.69,0,0.11}

\newcommand{\nicpb}{Laboratory of High Energy and Computational Physics, NICPB, R\"{a}vala pst. 10, 10143 Tallinn, Estonia}
\newcommand{\ippp}{Institute for Particle Physics Phenomenology, Department of Physics, Durham University, Durham DH1 3LE, United Kingdom}
\newcommand{\capfe}{CAFPE and Departamento de F\'isica Te\'orica y del Cosmos, Universidad de Granada, E-18071 Granada, Spain}

\begin{document}

\title{Confronting spin-3/2 and other new fermions with the muon g-2 measurement}

\author[a]{Juan C. Criado,}
\author[b,c]{Abdelhak Djouadi,}
\author[c]{Niko Koivunen,}
\author[c]{Kristjan M\"{u}\"{u}rsepp,} 
\author[c]{Martti Raidal,}
\author[c]{and Hardi~Veerm\"{a}e}  
\affiliation[a]{\ippp}  
\affiliation[b]{\capfe}  
\affiliation[c]{\nicpb} 

%
\emailAdd{juan.c.criado@durham.ac.uk}
\emailAdd{adjouadi@ugr.es}
\emailAdd{niko.koivunen@kbfi.ee}

\emailAdd{kristjan.muursepp@kbfi.ee}
\emailAdd{martti.raidal@cern.ch}
\emailAdd{hardi.veermae@cern.ch}

\abstract{
The new measurement of the muon's anomalous magnetic moment released by the Muon g-2 experiment at Fermilab sets strong constraints on the properties of many new particles. Using an effective field theory approach to the interactions of higher-spin fields, we evaluate the contribution of an electrically neutral and colour singlet spin-3/2 fermion to $(g-2)_\mu$ and derive the corresponding constraints on its mass and couplings. These constraints are then compared with the ones on spin-1/2 fermions, such as the vector-like leptons that are predicted by various extensions of the Standard Model, the excited leptons which appear in composite models, as well as the charginos and neutralinos of supersymmetric theories. Unlike these new spin-1/2 fermions, the spin-3/2 particles generate only small contributions to the muon anomalous magnetic moment. }
\maketitle

\section{Introduction}

The Muon $g-2$ collaboration at Fermilab has released~\cite{gm2-Fermilab,Abi:2021gix} the long awaited new measurement of the anomalous magnetic moment of the muon, $a_\mu= \frac12 (g-2)_\mu$. The result of the previous Brookhaven E821 muon $g-2$ experiment was~\cite{Bennett:2006fi} $a_\mu^{\rm E821}= 116 592 089(63)\times10^{-11}$, and had a deviation of about $3.7\sigma$, $\Delta a_\mu^{\rm E821}= a_\mu^{\rm E821} - a_\mu^{\rm SM} = (279 \pm 76)\times10^{-11}$ when compared with the recent worldwide consensus of the Standard Model (SM) contribution~\cite{Aoyama:2020ynm}
\be
\label{eq:g-2SM}
     a_\mu^{\rm SM}= (116 591 810 \pm 43)\times10^{-11},
\ee
giving rise to hopes for the presence of new physics contributions. The new result from the Fermilab experiment is~\cite{gm2-Fermilab,Abi:2021gix}
\be
\label{eq:g-2FL}
    a_\mu^{\rm FL}= (116 592 040 \pm 54)\times10^{-11},
\ee
which, by itself, represents a $3.3 \sigma$ deviation from the SM prediction.
When combined with the previous Brookhaven result one obtains~\cite{gm2-Fermilab,Abi:2021gix}
\be \label{eq:4sigma}
    a_\mu^{\rm EXP}= (116 592 061 \pm 41)\times10^{-11},
\ee
which implies a $4.2 \sigma$ deviation from the SM prediction,
$ \Delta a_\mu= a_\mu^{\rm EXP} - a_\mu^{\rm SM} = (251 \pm 59)\times10^{-11}$.
This result essentially confirms the previous Brookhaven $(g-2)_\mu$ measurement but with higher statistics. 
It is tempting to attribute the discrepancy $ \Delta a_\mu$ to new physics beyond the SM, but it could be due to still unknown theoretical and experimental uncertainties\footnote{A new lattice QCD analysis \cite{Borsanyi:2021} favors a result that is in accord with the SM prediction and is not included in the world average value that leads to the discrepancy with Eq.~(\ref{eq:4sigma}).}. In the latter case, the new measurement could serve to constrain new physics parameters in a way that is complementary to the intensive direct searches performed in the high-energy frontier at the Large Hadron Collider (LHC) experiments. 

In this paper, we confront the new $(g-2)_\mu$ result with the predictions coming from models beyond the SM, which contain new fermions. In particular, we consider the case of a massive electrically neutral and colourless spin-3/2 particle which was recently discussed~\cite{Criado:2020jkp,Criado:2021itq} but, to our knowledge, not in the context of lepton magnetic moments. We use the new measurement to derive constraints on the properties of these particles and compare these constraints with the existing ones on other spin-1/2 fermions appearing in various supersymmetric and non-supersymmetric theories beyond the SM.

Massive higher-spin particles, in particular spin-3/2 fermions in addition to spin-2 bosons are present in many supersymmetric extensions of gravity~\cite{Freedman:1976xh, Deser:1976eh, Freedman:1976py,VanNieuwenhuizen:1981ae}, and are also predicted in string-theoretical frameworks~\cite{Afkhami-Jeddi:2018apj,Kaplan:2020ldi}. Furthermore, they are of interest on purely phenomenological grounds as, for instance, they may constitute dark matter (DM)~\cite{Criado:2020jkp,Falkowski:2020fsu,Alexander:2020gmv,Gondolo:2020wge,Gondolo:2021fqo}, or appear in the form of non-conventional experimental signatures at the LHC and future colliders~\cite{Criado:2021itq}. As presently there is no well-motivated paradigm for searching for new phenomena beyond the SM, all possible directions must be pursued in the search for new physics and thus, considering the possibility of higher-spin fields is certainly worthwhile. 

In the past, phenomenological studies of generic higher-spin particles have had severe problems related to non-physical degrees of freedom in their representations which must be eliminated. For instance, spin-3/2 fermions described by Rarita-Schwinger fields~\cite{Rarita:1941mf} possess several pathologies, including causality and perturbative unitarity violation~\cite{Johnson:1960vt,Velo:1970ur}, unless identified with the supersymmetric partners of the graviton, the gravitino \cite{Grisaru:1977kk,Grisaru:1976vm}.

These difficulties are circumvented in the recently developed effective field theory (EFT) approach to generic higher-spin particles~\cite{Criado:2020jkp,Criado:2021itq}. This approach considers only the physical degrees of freedom of the higher-spin fields, and thus, although not admitting a Lagrangian description~\cite{Criado:2020jkp,Weinberg:1964cn}, allows for a consistent calculation of physical observables. It was successfully applied to study the DM~\cite{Criado:2020jkp} and high-energy collider phenomenology~\cite{Criado:2021itq} of electrically neutral and colour singlet higher-spin particles, with emphasis on spin-3/2 fermions and spin-2 bosons. While in this approach, the spin-2 particles couple only to gauge bosons, spin-3/2 particles have interactions with fermions, thus generating contributions to the lepton anomalous magnetic moments that have not been considered in the literature so far. This work fills this gap and provides the first example of a loop computation in this effective theory. 

For completeness, we will compare our results with models containing new spin-1/2 fermions, the existence of which is predicted by many extensions of the SM. Such states can have the usual lepton and baryon quantum numbers with exotic ${\rm SU(2)_L \times U(1)_Y}$ assignments. A prominent example is given by vector--like fermions, with both the left- and right-handed fermions appearing in the same electroweak multiplets, allowing for a consistent generation of their masses without invoking the Higgs mechanism. These often emerge in grand unified theories~\cite{Djouadi:1995es} and can be used to explain the SM  flavour hierarchies~\cite{Kannike:2011ng}. One can have sequential fermions, such as a fourth family, or mirror fermions that have chiral properties opposite to those of ordinary fermions. However, it is necessary to alter the Higgs sector of the SM in order to evade the stringent constraints from the measurements of the Higgs properties at the LHC~\cite{Djouadi:1997rj,Denner:2011vt,Djouadi:2012ae,Kuflik:2012ai}. The mixing of the new and the SM fermions with the same U(1)$_{\rm Q}$ and SU(3)$_{\rm C}$ assignments then gives rise to new interactions~\cite{Djouadi:1995es,Djouadi:1993pe}. These interactions allow for the decays of the heavy states into the lighter ones and to contributions that could be observed in high-precision experiments. 

Another type of new states are the excited fermions. Their appearance is a characteristic signature of substructure in the matter sector, which is used to explain patterns in the mass spectrum. Ordinary fermions may then correspond to the ground states of the spectrum containing excited states that can decay to the former via a magnetic type de-excitation. In the simplest scenarios, the excited fermions are assumed to have spin and isospin 1/2, and the transition between the excited and the fundamental states can be described by an ${\rm SU(3)_C \times SU(2)_L \times U(1)_Y}$ invariant effective interaction of the magnetic type~\cite{Djouadi:1995es,Boudjema:1992em}. Thus, besides the full couplings to gauge bosons, the excited states have couplings to the known fermions and gauge bosons which are inversely proportional to the compositeness scale $\Lambda$. These couplings determine the decay and production pattern of the excited states and induce contributions to the $(g-2)_\mu$ among other observables. 

Finally, we will very briefly discuss the case of supersymmetric theories (in their minimal version), in which the superpartners of the gauge and Higgs bosons, the spin-1/2 charginos and neutralinos, could also contribute to the $(g-2)_\mu$ along with their scalar partners, the smuons and their associated sneutrinos. These particles have long been sought for in order to provide an explanation to the previous discrepancy in the measurement.

The rest of the paper is organized as follows. In the next section, we present relevant details of the effective theory for a generic spin-3/2 fermion, perform the calculation for its contribution to the muon anomalous magnetic moment and confront the result with the new measurement. In section~\ref{spin-1/2}, we present the corresponding results for the spin-1/2 fermions: vector-like, excited and supersymmetric. We conclude in Section~\ref{conc}. Technical details relevant are presented in Appendix~\ref{app:loop_calc}.

\section{Muon's g-2 in a spin-3/2 effective framework}
\label{spin3/2}

\subsection{Spin-3/2 fermions in an effective field theory}

For a charge and colour neutral SM singlet spin-3/2 field in the Lorentz representation $(3/2,0)\otimes (0,3/2)$~\cite{Criado:2020jkp}, that we will denote by $\psi_{3/2}$, there are 6 independent linear dimension-7 operators that allow to describe its interaction with the SM fields~\cite{Criado:2021itq},
\bea\label{eq:L-linear}
	-\mathcal{H}_{\text{linear}} = & 	\frac{1}{\Lambda^3} \psi_{3/2}^{abc} \Big[
   c^{ijk}_q\epsilon^{IJK} u^{i\ast}_{RIa} d^{j\ast}_{RJb}d^{k\ast}_{RKc}  
+   c^{ijk}_l (L_{La}^{iT}\epsilon L^j_{Lb}) e^{k*}_{Rc}
+   c^{ijk}_{lq} (Q_{LIa}^{iT}\epsilon L_{Lb}^j)d_{Rc}^{kI\ast}  \\
& \!+\!	c_i^\phi \sigma^{\mu\nu}_{ab} (D_{\mu} \tilde{\phi})^\dagger  D_{\nu} L_{Lc}^i  
\!+\!	c_i^B \tilde{\phi}^\dagger \sigma^{\mu\nu}_{ab} B_{\mu\nu} L_{Lc}^i
\!+\!	c_i^W \tilde{\phi}^\dagger \sigma^{\mu\nu}_{ab} \sigma_n W^n_{\mu\nu} L_{Lc}^i 
\Big] 
   + \text{h.c.},
\eea
where $a,b,c$ are two-spinor indices, $I$ and $J$ are the colour indices, $i,j,k$ the flavour indices, and $n$ is the SU(2)-triplet index. The coefficient $c_q^{ijk}$ is symmetric in $jk$, while $c_l^{ijk}$ is symmetric in $ij$. $L^i_a$ and $Q^i_a$ are the left-handed lepton and quark doublets $L^i_{La} = (\nu_{La}^i, e_{La}^i)$ and $Q^i_{LIa} = (u_{LIa}^i, d_{LIa}^i)$, while $e_R^i$, $u_R^i$ and $d_R^i$ are the right-handed lepton and quark singlets. $B_{\mu \nu}$ and $W_{\mu \nu}$ denote the ${\rm U(1)_Y}$ and ${\rm SU(2)_L}$ field strengths and $\phi$ is the SM Higgs doublet. 
In the unitary gauge, $\phi = (0,H+v)/\sqrt{2}$, where $v$ is the vacuum expectation value $v=246$ GeV and $H$ the physical Higgs boson produced at the LHC ~\cite{Aad:2012tfa,Chatrchyan:2012ufa}. We define $D_{a\dot{a}} = \sigma^\mu_{a\dot{a}} D_\mu$, with $D_\mu$ being the usual 4-vector covariant derivative, and $\sigma^\mu_{a\dot{a}}$ given in terms of the identity matrix $\sigma^0$ and Pauli $\sigma^{1,2,3}$ matrices ${(\sigma^{\mu\nu})_a}^b \equiv i (\sigma^\mu_{a\dot{b}} \bar{\sigma}^{\nu\dot{b}b} - \sigma^\nu_{a\dot{b}} \bar{\sigma}^{\mu\dot{b}b})/4$. 

The advantage of this effective framework is that, unlike the Rarita-Schwinger field, the spin-3/2 field in Eq.~\eqref{eq:L-linear} carries only physical degrees of freedom~\cite{Criado:2020jkp}. This allows us to consistently study the physics of a generic massive spin-3/2 particle.

In this work, we concentrate on the muon anomalous magnetic moment for which the relevant effective Hamiltonian reads
\bea\label{eq:L}
	-\mathcal{H}_{\text{linear}} =	\frac{1}{\Lambda^3} \psi_{3/2}^{abc} \Big[ 
	& c_\mu (L_{La}^{2 T}\epsilon L^2_{Lb}) \mu^\ast_{Rc}
+	c_\phi \sigma^{\mu\nu}_{ab} (D_{\mu} \tilde{\phi})^\dagger D_{\nu} L_{Lc}^2 
   \\
+&	c_B \tilde{\phi}^\dagger \sigma^{\mu\nu}_{ab} B_{\mu\nu} L_{Lc}^2
+	c_W \tilde{\phi}^\dagger \sigma^{\mu\nu}_{ab} \sigma_n W^n_{\mu\nu} L_{Lc}^2 
\ \Big] 
   + \text{h.c.},
\eea
where we have kept only the $\psi_{3/2}$ couplings to the second generation leptons. In the following, we will set the coupling $c_\phi$ to zero because the corresponding operator does not contribute to the dipole moments. As we will show briefly, the contribution of $c_\mu$ operator to the electromagnetic dipole moments vanishes on-shell, and thus, the relevant phenomenology will be dictated by the last two operators in  Eq. \eqref{eq:L}. The Feynman rules for the various interactions of the spin-3/2 field, as well as the other details of our EFT of higher-spin, are listed in the appendices of Ref.~\cite{Criado:2021itq}.

\subsection{Estimating muon's g-2 from spin 3/2-fields }
\label{g-2-comp}

The anomalous magnetic and electric dipole moment (EDM) of the muon, $a_\mu \equiv \frac12 (g-2)_\mu$ and $d_\mu$ respectively, are defined through the amplitude~\cite{Raidal:2008jk}

\be
    \left<\mu(p)\right| A_\mu(q) \left|\mu(p + q)\right>
    =
    e F_1(q^2) \gamma_\mu
    + \frac{i e}{2 m_\mu} F_2(q^2) \sigma_{\mu\nu} q^\nu
    + \frac{i e}{2 m_\mu} F_3(q^2) \sigma_{\mu\nu} \gamma_5 q^\nu,
\ee
where $\left|\mu(p)\right>$ denotes the muon state with momentum $p$, $m_\mu$ is the muon mass, $e$ the electromagnetic coupling constant, and $\sigma_{\mu\nu}=\frac{i}{2} [\gamma_\mu,\gamma_\nu]$. Then, 
$a_\mu \equiv F_2(0)$, $d_\mu \equiv -\frac{e}{2 m_\mu} F_3(0)$ and $F_1(0) = 1$. 
In the low-energy EFT with only the SM fields, the leading order contribution to these form factors is generated by the operators
\be\label{g-2_4spinor}
    \mathcal{L}_{g-2, \rm EDM} =
    -\frac{e}{4 m_\mu} \hat{a}_\mu \, \bar \mu \sigma_{\mu\nu} \mu \, F^{\mu\nu}
    - \frac{i}{2} \hat{d}_\mu \, \bar \mu \sigma_{\mu\nu}\gamma_5 \mu \, F^{\mu\nu},
\ee
with $F^{\mu\nu}$ being the electromagnetic field strength tensor. In terms of their coefficients, we have
\be
    a_\mu = \hat{a}_\mu(\mu = m_\mu), \qquad
    d_\mu = \hat{d}_\mu(\mu = m_\mu),
\ee
where $\mu$ is the renormalization scale.

The dipole operators in Eq.~\eqref{g-2_4spinor} are written in terms of Dirac spinors, while the EFT formalism for spin-3/2 particle utilizes the 2-spinor formalism \cite{Dreiner:2008tw}. It is therefore convenient to write the dipole operators in terms of two-component spinors as
\bea\label{g-2_2spinor}
    \mathcal{L}_{g-2, \rm EDM} 
= & -\frac{1}{2}\left(\frac{e}{2 m_\mu} \hat{a}_\mu + i \hat{d}_\mu\right) ( \mu^{\dagger a}_R \sigma_{\mu\nu a}^{~~~~b} \mu_{L b})F^{\mu\nu}
    + \text{h.c.}\,.
\eea
One can see that the dipole operators are necessarily chirality changing.

\begin{figure}[t]
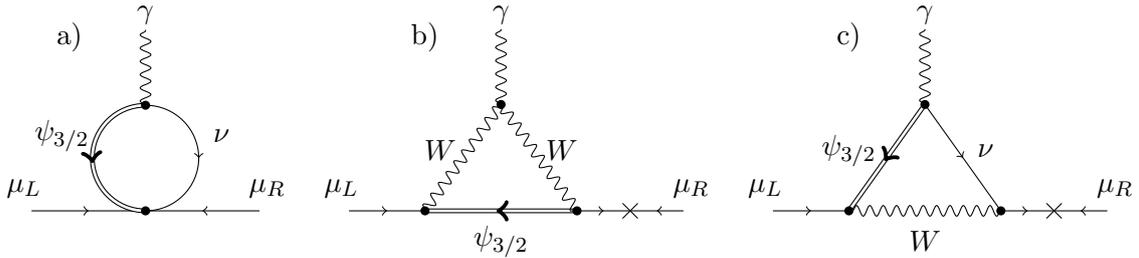

\centerline{  \diagramOne \diagramThree \diagramFive }
\caption[]{\label{fig:diagrams} Generic Feynman diagrams that can contribute to $g-2$ and the EDM of the muon. The crosses denote the chirality flip on the external leg.} 
\end{figure}

The diagrams that contribute to the anomalous magnetic moment must change the chirality of the incoming and outgoing muons. This chirality flip can take place inside the loop or on the external leg. At the one-loop level, there are three types of diagrams that can contribute to the dipole moments, depicted in Fig.~\ref{fig:diagrams} (the conjugate diagrams are not shown).

The contribution from diagram (a) in Fig.~\ref{fig:diagrams} is proportional to the square of the photon's momentum $q^2$ and thus vanishes on-shell.\footnote{By Lorentz invariance, the loop integral connecting the two vertices in Fig.~\ref{fig:diagrams}a can only be proportional to $q^{\dot a a}q^{\dot b b} \!+\! q^{\dot b a}q^{\dot a b}$ which implies that the induced form factors must be proportional to $q^2$.} Thus, only the diagrams (b) and (c) will contribute to muon's $g-2$ and EDM. They read
\bea
\label{eq:result}
    \hat{a}_\mu^{\psi}
&   = \frac{m_\mu^2 v^2 m_{3/2}^2}{8\pi^2\Lambda^6} \left[ |c_W|^2 f_1(m_{3/2}) + \frac{\textrm{Re} ( c_W^\ast c_{\gamma} )}{\sin(\theta_W)} f_2(m_{3/2}) \right],\\
    \hat{d}_\mu^{\psi}
&    = \frac{m_\mu v^2 m_{3/2}^2 g_2}{16\pi^2\Lambda^6} \textrm{Im} (c_W^\ast c_{\gamma}) f_2(m_{3/2}),
\eea
where $c_{\gamma} \equiv -c_B\cos\theta_W+c_W\sin\theta_W$ is the $\gamma\nu\psi_{3/2}$ coupling, $\theta_W$ is the Weinberg angle and the functions $f_1$, $f_2$ are given in Appendix \ref{app:loop_calc}. The loops are divergent, and we use dimensional regularization to regularize the integrals. When $m_{3/2} \gg m_W$, the $\overline{MS}$-scheme renormalized contributions are approximately
\bea
    f_1 = -\frac{13}{27} + \frac{7}{18}\log\left(\frac{\mu^2}{m_{3/2}^2}\right), \qquad
    f_2 = \frac{2}{3} \log\left(\frac{\mu^2}{m_{3/2}^2}\right),
\eea
where $\mu$ is the renormalization scale. 

We remark that the contribution to the muon magnetic moment in Eq.~\eqref{eq:result} is of the order $\Lambda^{-6}$ while, by naive power counting arguments, the UV-physics responsible for the effective Hamiltonian~\eqref{eq:L} could generate the dimension-5 SM dipole operators~\eqref{g-2_4spinor} already at the order $\Lambda^{-1}$ in the EFT. However, depending on the UV completion, the dipole operators and $\psi_{3/2}$ interactions may be generated at very different scales. For example, if $\psi_{3/2}$ is composite, then $\Lambda$ may be associated with its compositeness scale. If the UV-physics responsible for a composite $\psi_{3/2}$ does not generate magnetic moments directly, then the latter will be suppressed by some scale higher than $\Lambda$. So, the contribution~\eqref{eq:result} from the loops in Fig.~\ref{fig:diagrams} can be the leading one.

The magnetic moment in Eq.~\eqref{eq:result} is evaluated at some high scale. To run its value down to the muon mass scale, we closely follow Ref.~\cite{Aebischer:2021uvt}, in which the running and matching from several scales to low energies relevant for the muon dipole moments has been computed. We assume that $m_{3/2}$ is sufficiently close to $\SI{250}{GeV}$, so that we can fix the renormalization scale $\mu$ to this value. Then,
\be
    a_\mu^\psi = 0.89 \, \hat{a}_\mu^\psi(\mu = \SI{250}{GeV}).
\ee

\begin{figure}[t]
\centering
 \hspace{-4mm}
 \includegraphics[width=1.0\linewidth]{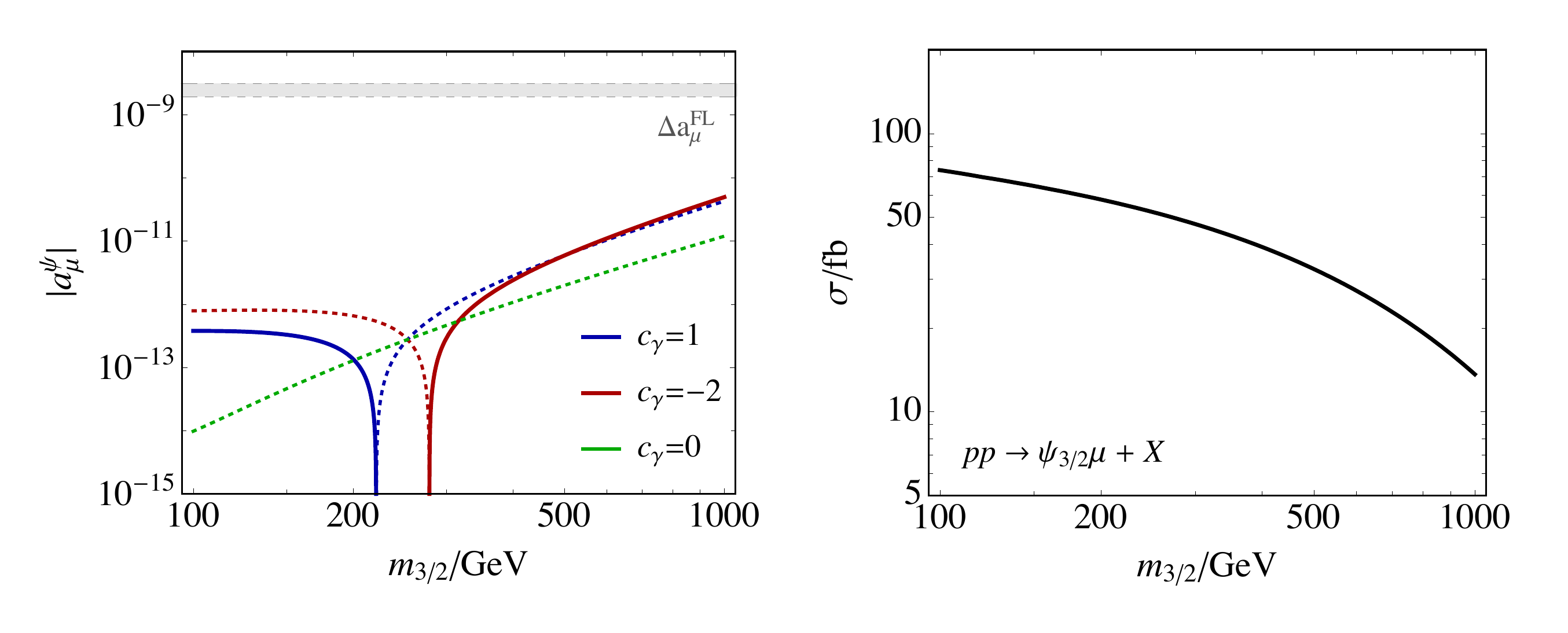} 
\caption[]{{\it Left:} $a_\mu^\psi$ for different $m_{3/2}$ and $c_{\gamma}$ values while the other parameters are fixed as $\Lambda=1$~TeV, $\mu=250$~GeV, $c_W=1$ and $c_\phi=0$. The solid and dashed lines correspond to $a_\mu^\psi>0$ and $a_\mu^\psi<0$, respectively. The light grey region represents the $1\sigma$ band of the Fermilab measurement. {\it Right:} The production cross section of $\psi_{3/2}$ particles at the LHC as a function $m_{3/2}$ in the process $pp \to \psi_{3/2} \mu +X$ at $\sqrt s= 14$ TeV. The production cross section is independent of $c_{\gamma}$.}
\label{fig2}
\end{figure}

The contribution to $(g-2)_{\mu}$ from the higher-spin field for different values of $m_{3/2}$ and $c_{\gamma}$ is shown in the left-panel of Fig.~\ref{fig2}. The results can be roughly summarized as
\be\label{a_mu_bound}
    |a_\mu^\psi| \lesssim 2 \times 10^{-11} \left[ \frac{\Lambda}{\TeV} \right]^{-6} \left[ \frac{m_{3/2}}{\TeV} \right]^{2},
\ee
when $c_W, c_\gamma<1$ as expected in the EFT approach. This contribution to $(g-2)_{\mu}$ is consistent with the SM unless the EFT scale is close to the EW scale, $\Lambda \lesssim 250 \GeV$, in which case the validity of the EFT approach is questionable. Also, note that Fig.~\ref{fig2} mildly violates Eq.~\eqref{a_mu_bound} for masses close to the EFT scale. This behaviour is just an artefact of the large logarithm $\log(m_{3/2}/\mu)$. Moreover, $a_\mu^\psi$ is negative when $c_\gamma = 0$. A positive $a_\mu^\psi$ may be obtained for specific values of the model parameters, that is, for sufficiently low $m_{3/2}$ when $c_\gamma > 0$, or for high enough $m_{3/2}$ when $c_\gamma < 0$.

As can be seen, the spin-3/2 contribution to $(g-2)_{\mu}$ is typically of order $10^{-11}$, more than an order of magnitude below the experimental sensitivity in the most favourable case. Nevertheless, for a particle with such mass and couplings, the production cross section at the LHC in the process $pp \to q\bar q' \to W^*\to  \psi_{3/2} \mu +X$, as calculated in Ref.~\cite{Criado:2021itq}, is shown in the right panel of Fig.~\ref{fig2}. It is rather significant, reaching a level of 10~fb at very high masses, a rate that should be sufficient to observe the particle (which could mainly decay into a clear signature consisting, \eg, of a $W$ boson and a muon) at the next LHC runs with expected integrated luminosities of several 100~fb$^{-1}$ to several ab$^{-1}$.

Finally, $\psi_{3/2}$ may induce an EDM for the electron and the muon. These are subject to the following bounds~\cite{Zyla:2020zbs}
\bea
    d_e &< 0.87\times 10^{-28}~ \rm{e}~ \rm{cm},\qquad d_\mu &< 1.8\times 10^{-19}~ \rm{e}~ \rm{cm}, 
\eea
which translate into bounds on the imaginary parts of the couplings $c_W$ and $c_\gamma$ as
\bea
    &\textrm{Im} \left[c_W^{e\ast} c^e_\gamma\right] < 1.6\times 10^{-14}~ \log\left(\frac{\mu^2}{m^2_{3/2}}\right) \frac{\Lambda^6}{m_{3/2}^2 \textrm{GeV}^4},\\
    &\textrm{Im} \left[c_W^{\mu\ast} c^\mu_\gamma\right] < 1.7\times 10^{-8}~ \log\left(\frac{\mu^2}{m^2_{3/2}}\right) \frac{\Lambda^6}{m_{3/2}^2 \textrm{GeV}^4}.
\eea 
In particular, for the choice of parameters used above, $\mu=250$ GeV, $m_{3/2}=200$ GeV and $\Lambda=1000$ GeV, the constraints read
\bea
    &\textrm{Im} \left[c_W^{e\ast} c^e_\gamma\right] < 0.88,  \qquad
    &\textrm{Im} \left[c_W^{\mu\ast} c^\mu_\gamma\right] < 8.8\times 10^{6}.
\eea
Thus, current observations easily allow for $\mathcal{O}(1)$ imaginary parts of $c_W^{\mu}$, $c^\mu_\gamma$ but can be mildly constraining in the case of the electron.

\section{Constraints on new spin-1/2 fermions}
\label{spin-1/2}

\subsection{Vector-like leptons}

For charged heavy leptons with exotic ${\rm SU(2)_L \times U(1)_Y}$ quantum numbers, except for singlet neutrinos without electromagnetic or weak charges, the couplings to the photon, to the $W$ and the $Z$ bosons are unsuppressed. The new states mix with the SM leptons in a model-dependent and a possibly rather complicated way, especially if different generations can mix. 

In the following, we will consider the case of vector-like leptons that have been introduced in Refs.~\cite{Kannike:2011ng,Giudice:2012ms} in order to explain flavour hierarchies in the SM. Two doublets $L_L,L_R$ and two singlets $E_L, E_R$ are introduced with the Lagrangian
\be \label{eq:lagrangian}
 {\cal L} \! \propto \! M_E \bar{E}_{L} E_{R} \!+\! M_L \bar{L}_{R} L_{L} \!+\! m_E \bar{E}_{L} e_{R} \!+\!
m_L \bar{L}_{R} \ell_{L} \!+\! \lambda_{LE} \bar{L}_{L} E_{R} \Phi \!+\! {\bar\lambda}_{LE} \bar{L}_{R} E_{L} \Phi^\dagger
\!+\! {\rm h.c.},
\ee
with $L_L, E_R$ and $L_R,E_L$ states having, respectively, the same and opposite quantum numbers as the SM leptons $\ell_L, e_R$; $\Phi$ is the SM Higgs doublet defined before. The mass eigenstates are obtained by diagonalizing the mass mixing in ${\cal L}$ through $2 \times 2$ unitary matrices, where the mixing angles are $\tan\theta_L = m_L/M_L$ and $\tan\theta_R = m_E/M_E$. Upon rotating the fields, the previous Lagrangian becomes 
\bea
    {\cal L} & \propto \sqrt{M^2_L + m^2_L}\, \bar{E}_{L} E_{R} 
    + \sqrt{M^2_E + m^2_E}\, \bar{L}_{R} L_{L} + {\bar\lambda}_{LE}\, \bar{L}_R E_L \Phi^\dagger \nonumber \\
    &+ \lambda_{LE}\, \left( s_{\theta_L} s_{\theta_R} \bar{\ell}_L e_R  \!+\! 
    c_{\theta_L} s_{\theta_R} \bar{L}_L e_R \!+\! s_{\theta_L} c_{\theta_R} \bar{\ell}_LE_R \! +\! c_{\theta_L} c_R~\bar{L}_L E_R \right) \Phi + {\rm h.c.}\,,
\eea
where $s_\theta \equiv \sin\theta$ and $c_\theta \equiv \cos\theta$. After spontaneous symmetry breaking, the spectrum will consist of two heavy leptons with masses $\sqrt{M^2_L + m^2_L}$ and $\sqrt{M^2_E + m^2_E}$ and light leptons with masses given approximately by $m_{\ell_i} \simeq \lambda_{LE}\, s^i_{\theta_L} s^i_{\theta_R} v+ \mathcal{O}(v^2/ M^2_{L,E})$, where $i$ is the generation index.

The heavy charged and neutral leptons contribute to the anomalous magnetic moment through Feynman diagrams that are similar to those of (b) and (c) depicted in Fig.~\ref{fig:diagrams}. They involve the exchange of two $W$ bosons with the neutral lepton and the exchange of two charged states with a $Z$ or Higgs boson. Heavy exotic fermion contributions to leptonic $(g-2)$ have been discussed in Refs.~\cite{Rizzo:1985db,Nardi:1991rg,Vendramin:1988zc,Kephart:2001iu,Dermisek:2013gta,Csikor:1991np,Chavez:2006he,Aboubrahim:2016xuz}. In our case, they have been evaluated and analytical expressions have been given in Ref.~\cite{Kannike:2011ng} to which we refer for the details. Here, we simply give the contributions to $a_\mu$ in the limit of small mixing angles, retaining only terms of order $v^2 / M^2_{L,E}$. In this case, one finds \cite{Kannike:2011ng}
\be
    \Delta a_\mu \simeq \frac{1}{16\pi^2}\frac{m^2_\mu}{M_L M_E} {\rm Re}(\lambda_{LE}{\bar\lambda_{LE}}) \approx  10^{-9}~{\rm Re}(\lambda_{LE}{\bar\lambda_{LE}})\,
\left(\frac{300~{\rm GeV}}{\sqrt{M_L M_E}}\right)^2.
\ee
Hence, for $M_L,M_E$ of the order of the electroweak symmetry breaking scale $v$ and for large Yukawa couplings to the muon $\lambda_{LE}, \bar\lambda_{LE}$, the contributions to $a_\mu$ can be significant.

\subsection{Excited leptons}

In the case of the excited leptons that we will denote by $\ell^\star$, we assume for simplicity that they have spin and isospin 1/2. Besides the $\ell^\star \ell^\star V$ interaction with the $V=\gamma, W, Z$, gauge bosons, there is a magnetic-type coupling between the excited leptons, the ordinary leptons and the gauge bosons $\ell^\star \ell V$ which allows for the decays of the heavy states, $\ell^\star \to V\ell$ \cite{Boudjema:1992em}. This coupling induces a contribution to the anomalous magnetic moment of the lepton via diagrams similar to (b) and (c) in Fig.~\ref{fig:diagrams} (upon replacing the $\psi_{3/2} \nu W$ loop by the $\mu^\star \mu^\star$ loop along with the $Z$ boson and the photon, as well as diagrams in which the magnetic transition occurs at the $\gamma \mu^\star \mu$ vertex). The Lagrangian describing this transition should respect a chiral symmetry in order to not induce excessively large contributions to these moments. Consequently, only the left- or the right-handed component of the excited lepton takes part in the generalized magnetic de-excitation. The Lagrangian thus reads 
\be
    {\cal L}_{ \ell \ell^{\star} \gamma} 
    = \frac{e \kappa_{L/R} }{\sqrt 2 \Lambda} \bar{\ell^\star} \sigma^{\mu \nu} \ell_{L/R} F_{\mu \nu}+ {\rm h.c.}\,.
\ee
This could be generalized to the ${\rm SU(2)_L \times U_Y(1)}$ case where the photon field
strength is extended to the $W_{\mu\nu}$ and $B_{\mu \nu}$ ones. In the equation above, 
$\Lambda$ is the compositeness scale that we equate to 1 TeV. We will set all the weight factors for the field strengths to $\kappa_{L/R}$ to simplify the analysis and to ensure that the excited neutrino has no tree-level electromagnetic couplings~\cite{Boudjema:1992em}. Thus, apart from the masses of the excited leptons that we will also equate: $m_{\ell^\star}= m_{\nu^\star_\ell}$, the only free parameter will be the strength $\kappa_{L,R}/\Lambda$ of the de-excitation that involves either a left-handed or a right-handed fermion.

The contribution $a_\mu$ of the excited muon $\mu^\star$ and its neutrino partner $\nu^\star_\mu$ to the muon magnetic moment has been calculated long ago \cite{Renard:1982ij,delAguila:1984sw,Choudhury:1984bu,Mery:1989dx,Rakshit:2001xs} and the result in the case where the simplifications above are made, assuming $m_{\ell^\star}= m_{\nu^\star_\ell}= \Lambda \gg m_W$, which we anticipate to be a good approximation, is simply given by 
\cite{Mery:1989dx}
\be
    \Delta a_\mu = \frac{\alpha}{\pi} \frac{\kappa_{L,R}^2}{\Lambda^2} m_\mu^2 c_{L/R},
\ee
where the numerical values in these limits are $c_L \simeq 10$ and $c_R \simeq 5.3$, respectively for left-handed $V \mu^\star \mu_L$ and right-handed $V \mu^\star \mu_R$ transitions. 

\subsection{Supersymmetric particles}
For the sake of completeness, we will briefly discuss the contributions to $a_\mu$ of the superparticles in the minimal supersymmetric extension of the SM \cite{Drees:2004jm}, namely the one with the chargino-sneutrino and neutralino-smuon loops. These have been calculated by several authors \cite{Ellis:1982by,Grifols:1982vx,Barbieri:1982aj,Kosower:1983yw, Chattopadhyay:2001vx,Carena:1996qa,Martin:2001st,Chakraborti:2021kkr,Moroi:1995yh} and the approximate result, taking into account only the chargino-sneutrino loop contribution which is an order of magnitude larger than that of the neutralino-smuon loop, is rather simple and accurate\footnote{Note that the sign of the SUSY contribution is equal to the sign of the higgsino parameter $\mu$, $\Delta a_\mu \propto (\alpha/\pi) \times \tan\beta (\mu M_2)/ \tilde m^4$ with $M_2$ the gaugino (wino) mass parameter.}~\cite{Moroi:1995yh} 
\be
    \Delta a_\mu 
    \simeq \frac{\alpha}{8 \pi s_W^2} \tan\beta \times \frac{m_\mu^2} {\tilde m^2} 
    \approx 1.5 \times 10^{-11} \tan\beta \left[ \frac{\tilde m}{\TeV} \right]^{-2},
\ee
where $\tan\beta$ is the ratio of vacuum expectation values of the two doublet Higgs fields that break the electroweak symmetry, $1 \lsim \tan\beta \lsim m_t/m_b \approx 60$ and $\tilde m$ is a supersymmetric scale given by the largest mass among the chargino and the sneutrino states $\tilde m = {\rm max} (m_{\tilde \nu}, m_{\chi_1^+})$. Thus, a large SUSY contribution to $a_\mu$ can be generated for high enough $\tan\beta$ values and small chargino and second generation slepton masses, of the order of few hundred GeV. 

\subsection{Constraints}

\begin{figure}[t]
\centering
 \hspace{-8mm}
 \includegraphics[width=0.65\linewidth]{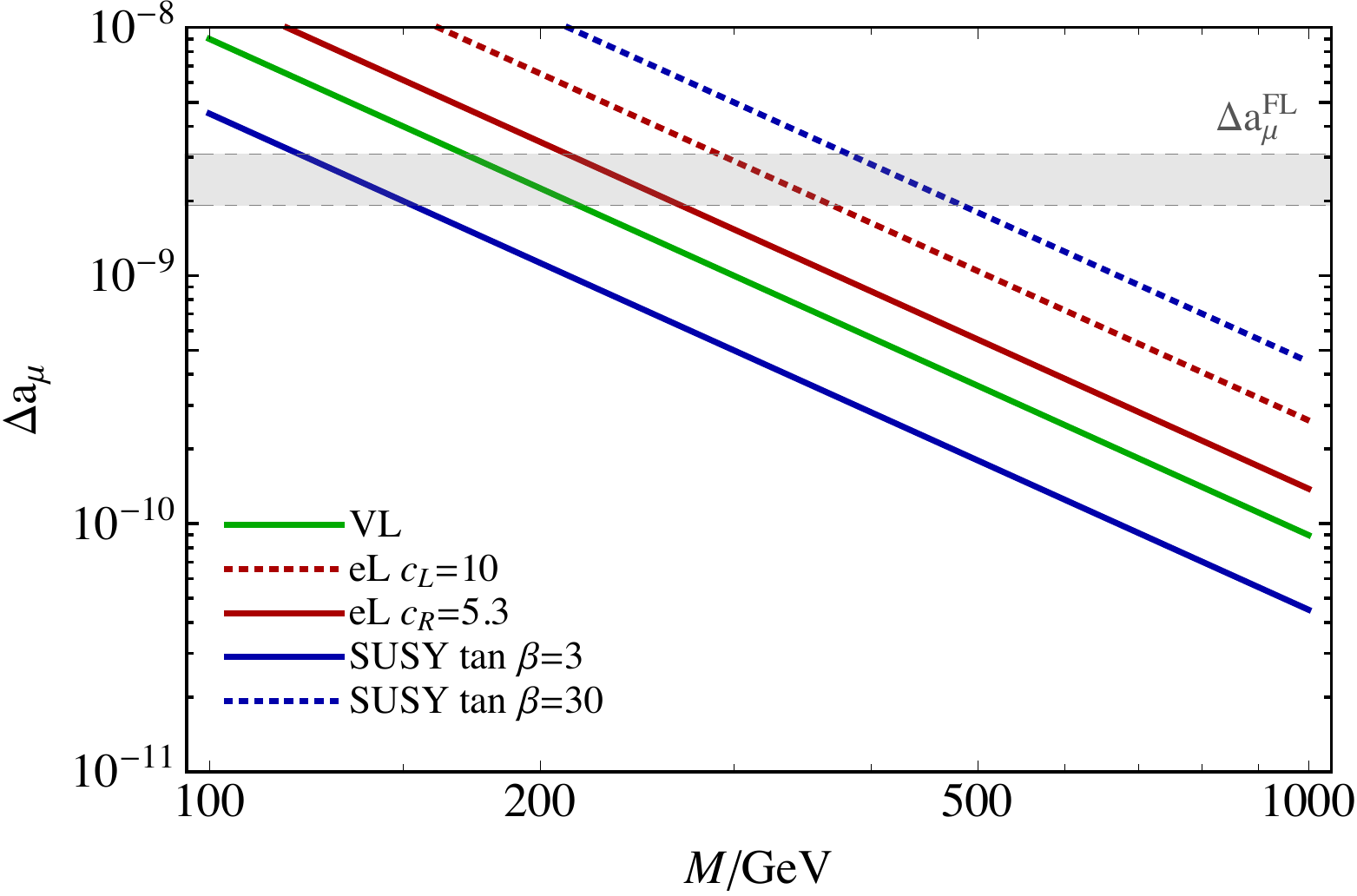}
\caption[]{
Contributions to the $(g-2)_\mu$ from various spin-1/2 particles as functions of the relevant mass scale $M$: vector-like leptons for $\lambda_{LE} = \overline{\lambda}_{LE} = 1$ and $M = \sqrt{M_{L}M_{E}}$ (green line), excited leptons with $\kappa_{L,R}= 1$ and $M=\Lambda$ for the two cases $c_{L} \approx 10$ (dotted red line) and $c_{R} \approx 5.3$ (solid red line) and supersymmetric particles for $M=\tilde m$ for the two cases of $\tan \beta = 3$ (solid blue line) and $\tan \beta = 30$ (dotted blue line). The light grey band shows the $1\sigma$ region of the Fermilab measurement.
}
\label{fig3}
\end{figure} 

Our numerical results for the three cases of exotic spin-1/2 fermions discussed in the previous subsections are collected in Fig.~\ref{fig3} where we present the typical predictions for their contributions to $(g-2)_\mu$ as functions of their representative mass scales. In the case of vector-like fermions, the $M$ scale is defined as $M = \sqrt{M_{L}M_{E}}$ which, along with the assumption that the Yukawa couplings are given by $\lambda_{LE} = \overline{\lambda}_{LE} = 1$, leads to the curve depicted in green in Fig.~\ref{fig3}. For excited leptons, the scale $M$ is defined as $M=\Lambda= m_{\mu^*} = m_{\nu_\mu^*}$, and we have considered two situations, $\kappa_L=1$ and $\kappa_R=1$, which lead to the coefficients $c_L=10$ and $c_R=5.3$ respectively. The resulting contributions to $a_\mu$ are presented in the figure in red. Finally, in the supersymmetric case, the scale is simply $M= \tilde m$, and we have chosen $\tan\beta=3$ and $\tan\beta=30$ to illustrate our results. The resulting typical contributions to $a_\mu$ are shown by the blue curves in Fig.~\ref{fig3}. The results of the new Fermilab $(g-2)_\mu$ measurement at $1\sigma$ are presented by the grey band. 

Comparison of the predicted results with the new $(g-2)_\mu$ measurement indicates that all the considered spin-1/2 scenarios could explain the discrepancy with respect to the SM prediction. In turn,  if the latter discrepancy is to be attributed to additional theoretical errors for instance, the models would be severely constrained by the experiment and,  typically, the scale of new physics would be constrained to be above several hundred GeV. This is in deep contrast with the spin-3/2 case which predicts $\Delta a_\mu \sim 10^{-11}$ and, therefore, direct searches at colliders would remain the most powerful tool to search for higher-spin particles.

\section{Discussion and conclusions}
\label{conc}

In this work, we have computed the contribution of a generic massive SM singlet spin-3/2 fermion to the muon anomalous magnetic moment $(g-2)_\mu$ and to the muon electric dipole moment $d_\mu$. We have used an effective field theory to describe the higher-spin fermion interactions which involve only the physical degrees of freedom, allowing us to compute physical observables in a consistent way. This is in sharp contrast with the situation of the Rarita-Schwinger spin-3/2 field, for which interactions can excite unphysical degrees of freedom unless the field is identified with the gravitino of supergravity. 

Our results show that the spin-3/2 induced left-right chirality flipping operators to the $(g-2)_\mu$. In particular, the ones that are expected to lead to enhanced contributions to $(g-2)_\mu$ vanish in the limit of an on-shell photon, $q^2=0$. The leading non-vanishing contributions, therefore, are suppressed by the chirality flip on the external muon leg and cannot compete with the SM contributions. As a consequence, spin-3/2 particles 
even for masses close to the electroweak scale cannot explain the new experimental result from Fermilab in Eq.~\eqref{eq:g-2FL} if the discrepancy with the SM result is indeed real. If not, the new result will not impose significant constraints on the properties of the spin-3/2 particle and would leave open the possibility of producing such particles at the next LHC runs. The corresponding numerical results are presented in Fig.~\ref{fig2} which reports the predicted values for $(g-2)_\mu$ as well as the corresponding $\psi_{3/2}$ production cross section at the LHC. The latter can reach values of order 100~fb. Therefore, direct searches at colliders would remain the most powerful tool for discovering the higher-spin particles.

For the sake of comparison and completeness, we have briefly considered various hypothetical spin-1/2 fermion contributions to the $(g-2)_\mu$, such as the ones of new leptons with exotic ${\rm SU(2)_L \times U(1)_Y}$ quantum numbers, excited leptons of composite models and supersymmetric particles, namely the combined contributions of neutralinos/charginos with smuons and their associated sneutrinos. Our results for the exotic spin-1/2 fermions are summarised in Fig.~\ref{fig3}. All these particles give a much larger contribution to $(g-2)_\mu$ which, when confronted with the latest experimental measurement, imply masses below the TeV scale for these states, should they explain the discrepancy from the SM expectation. 

\vspace{5mm}
\noindent \textbf{Acknowledgement:} This work was supported by the Estonian Research Council grants MOBTTP135, PRG803, MOBTT5, MOBJD323 and MOBTT86, and by the EU through the European Regional Development Fund CoE program TK133 ``The Dark Side of the Universe." J.C.C. is supported by the STFC under grant ST/P001246/1, and A.D. is also supported by the Junta de Andalucia through the Talentia Senior program as well as by A-FQM-211-UGR18, P18-FR-4314 with ERDF.

\appendix
\section{Spin-3/2 loop functions}
\label{app:loop_calc}

We use dimensional regularization and the $\overline{\rm{MS}}$--renormalization scheme to extract the finite component of the loop integrals. All expressions are given up to the leading term in $m_\mu$. 

The $WW\psi_{3/2}$ loop gives
\bea
    a_\mu^{WW\psi} 
&   =  \frac{m_\mu^2 v^2 m_{3/2}^2}{8 \pi^2 \Lambda^6} |c_W|^2 f_1(m_{3/2}),
    \qquad
    d_\mu^{WW\psi}
   = 0,
\eea
where
\bea
    f_1(m)
    & = \int^1_0 \td x \frac{\Delta}{6 m^2}
    \left[-10+5x+7x - 4x(11-15x)\ln(\Delta/\mu^2))\right] \\
    &= \frac{1}{108 m^2 (m^2-m_W^2)^3}  \Bigg[-52m^8+55m^6 m_W^2 -36m^4 m_W^4 +25m^2 m_W^6 +8m_W^8 \\
    &+  6m^6(-7 m^2 + 22 m_W^2)\log\left(\frac{m^2}{m_W^2}\right) +6(m^2-m_W^2)^3(7m^2-m_W^2)\log\left(\frac{\mu^2}{m_W^2}\right)\Bigg],
\eea   
and $\Delta = m_{3/2}^2 + x^2 m_\mu^2-(m_{3/2}^2+m_\mu^2-m_W^2)x$. In the limiting case $m_{3/2}\gg m_W$, we have that
\bea
    f_1(m) = -\frac{13}{26} + \frac{7}{18}\log\left(\frac{\mu^2}{m^2}\right) + \mathcal{O}(m_W^2/m^2).
\eea

The $W\nu\psi_{3/2}$ loop gives
\bea
    a_\mu^{\psi\nu W} 
&   = \frac{ m_\mu^2 v^2 m_{3/2}^2 }{8 \pi^2 \Lambda^6 \sin(\theta_W)} \textrm{Re} \left[c_W^\ast c_\gamma\right]f_2(m_{3/2}),\\
    d_\mu^{\psi\nu W} 
&    =  \frac{m_\mu v^2 m_{3/2}^2 g_2 }{16 \pi^2 \Lambda^6 } \textrm{Im} 
    \left[c_W^\ast c_{\gamma} \right] f_2(m_{3/2}),
\eea
where $c_{\gamma} \equiv -c_B\cos\theta_W+c_W\sin\theta_W$,
\bea
    f_2(m) 
    & = \int^1_0 \td x\int^x_0 \td y  \frac{4\Delta}{m_{3/2}}
    \left[-1-y/3 - 4y\ln(\Delta/\mu^2)\right] \\
    & =   \frac{2}{3 m^2 (m^2-m_W^2)^2} \Bigg[m^4 m_W^2 -m^2 m_W^4 -  m^6\log\left(\frac{m^2}{m_W^2}\right)\\
    &\qquad\qquad\qquad\qquad\qquad\qquad +(m^2-m_W^2)^2(m^2+2m_W^2)\log\left(\frac{\mu^2}{m_W^2}\right)\Bigg],
\eea 
and $\Delta = m_\mu^2y^2 + m_{3/2}^2(1-x) + (m_W^2-m_\mu^2)y$. In the limiting case $m_{3/2}\gg m_W$,
\bea
    f_2(m) = \frac{2}{3} \log\left(\frac{\mu^2}{m^2}\right) + \mathcal{O}(m_W^2/m^2).
\eea

\bibliography{any_spin}

\end{document}